\documentstyle[11pt,newpasp,twoside,epsf]{article}
\reversemarginpar

\begin{document}
\title{Stellar Coronae with \textit{XMM-Newton} RGS\\
 II. X-ray Variability}

\author{M. Audard, M. G\"udel}
\affil{Paul Scherrer Institut, W\"urenlingen \& Villigen, 5232 Villigen PSI,
       Switzerland}
\author{A.~J. den Boggende, A.~C. Brinkman, J.~W. den Herder, J.~S. Kaastra,
        R. Mewe, A.~J.~J. Raassen, C. de Vries}
\affil{Space Research Organization of the Netherlands, Sorbonnelaan 2, 3584 CA Utrecht,
       The Netherlands}
\author{E. Behar, J. Cottam, S.~M. Kahn, F.~B.~S. Paerels, J.~M. Peterson, 
        A.~P. Rasmussen, M. Sako}
\affil{Columbia Astrophysics Laboratory, Columbia University, 550 West 120th Street, New York, 
        NY 10027, USA}
\author{G. Branduardi-Raymont, I. Sakelliou}
\affil{Mullard Space Science Laboratory, University College London, Holmbury St.~Mary,
 Dorking, Surrey, RH5 6NT, United Kingdom}
\author{C. Erd}
\affil{Astrophysics Division, Space Science Department of ESA, ESTEC, 2200 AG Noordwijk, 
       The Netherlands}

\begin{abstract}
First results from high-resolution coronal spectroscopy of flares with the
Reflection Grating Spectrometers on board the \textit{XMM-Newton} satellite 
are reviewed. Rotational modulation in the X-ray light curve of HR 1099 is 
discussed.
Results from time-dependent spectroscopy of flares in the active 
stars HR 1099, AB Dor, YY Gem are also presented. Variations in the shape of 
the emission measure distributions, in the abundances and in the average 
density of the cool plasma are discussed.
\end{abstract}
%___________________________________________________________________________
\section{Introduction}
Stellar coronae often display variability in their X-ray emission. Energetic
explosive events (flares), eclipses, and rotational modulation are among the
most interesting features found in X-ray light curves of coronal stars. 
Flares are at the center of a scenario that assumes an ensemble of flares as 
agents of heating of stellar coronae (``microflare hypothesis''; e.g., 
Parker~1988). Recent results suggest that flares can contribute significantly 
to the coronal heating of active stars (Audard et al.~2000). In a standard 
model, magnetic energy is released during a flare, heating the dense layers of 
chromospheric material. Hot ionized material is then driven into the corona by 
the pressure increase and fills the coronal loops (chromospheric evaporation; 
see Antonucci et al.~1994).  Optically thin X-ray radiation occurs through 
continuum processes  and electronic transitions.  Flares are therefore 
important agents that bring ``fresh'' material from the chromospheric layers 
to the corona. 

In the non-flaring Sun, a First Ionization
Potential (FIP) effect is observed, with enhanced coronal abundances for elements with
low FIP ($<10$~eV), while high-FIP elements show solar photospheric values.
In the stellar case, evidence for the presence of the FIP effect is less clear,
but has been found
in some stars (e.g., Drake et al.~1995; 1997). Recently, Brinkman et al.~(2001) 
has found in HR 1099 evidence for an \emph{inverse} FIP effect and
a strong enhancement of Ne relative to O. Stellar coronal flares often show
enhanced metal abundances (e.g., Ottmann \& Schmitt~1996). 
Also, some classes of solar flares can be Ca-rich (Ca is a low-FIP
element; Sylwester et al.~1984) or Ne and S-rich (Ne and S are high-FIP
elements; Schmelz 1993). 

In this paper, we report a summary of first results (see also G\"udel et al.,
this volume) from studies of stellar flares with the \textit{XMM-Newton} 
Reflection Grating Spectrometers (RGS). Recent observations include HR 1099, 
AB Dor, YY Gem/Castor  (see Audard, G\"udel, \& Mewe~2001 and G\"udel et 
al.~2001ab for more details).
%___________________________________________________________________________
\section{X-ray rotational modulation in HR 1099}
The RGS observed the RS CVn binary star HR 1099 (V711 Tau) for a time span of about 25 days (Fig.~1, left).
The light curve shows variability on short and long
time scales. Similar to previous results (Agrawal \& Vaidya~1988; Drake et al.~1994), 
rotational modulation is suggested, with maximum flux when the
more active star (K1 IV) is in front. Recently, Ayres et al.~(2001) found 
consistency between line wavelength shifts and the orbital motion of the K 
component of HR 1099. Taken together, these observations
suggest that one or several active regions are present on the hemisphere of the
K star that is facing away from its companion (Audard et al.~2001). 
\begin{figure}[!t]
\plottwo{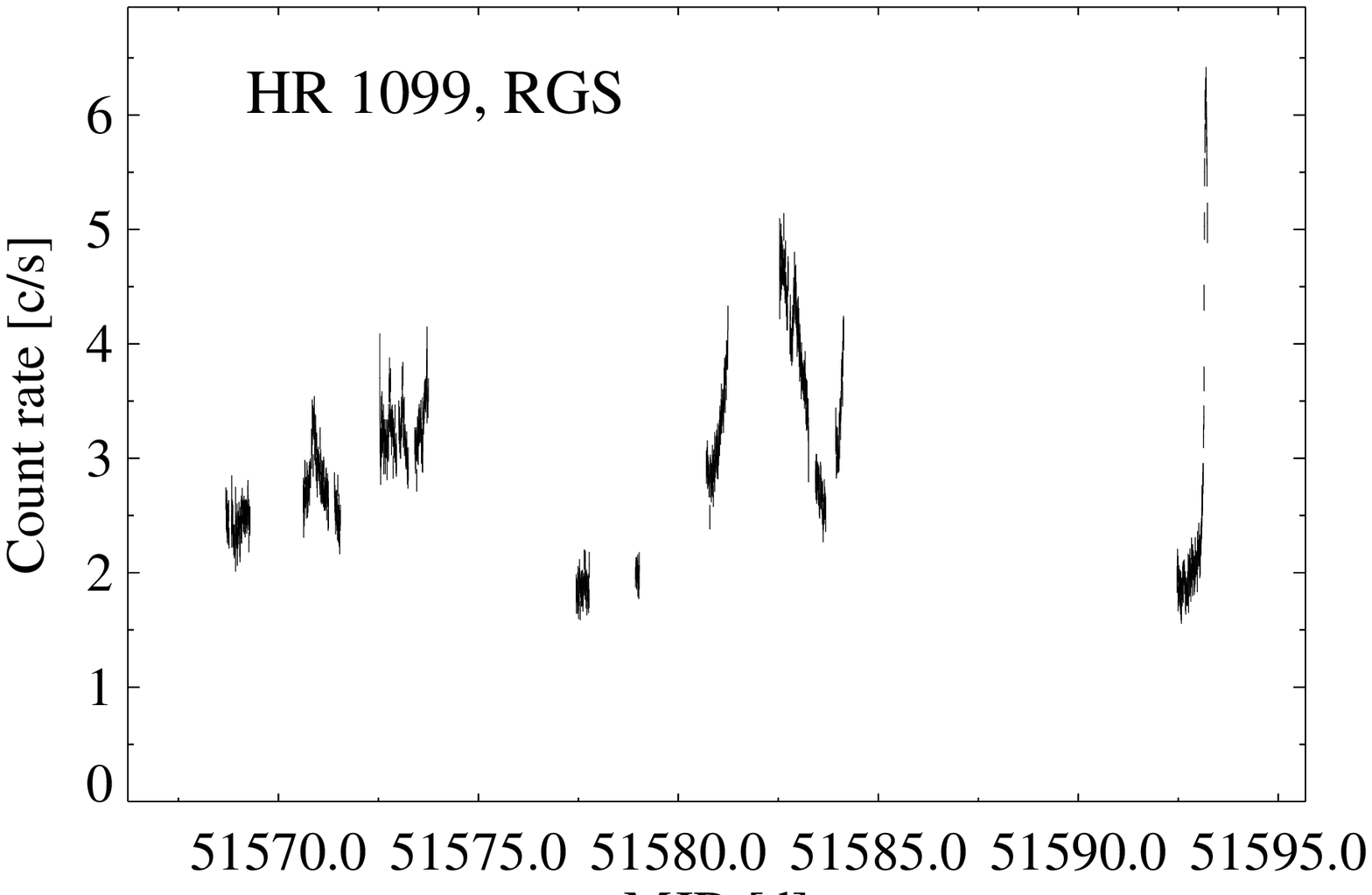}{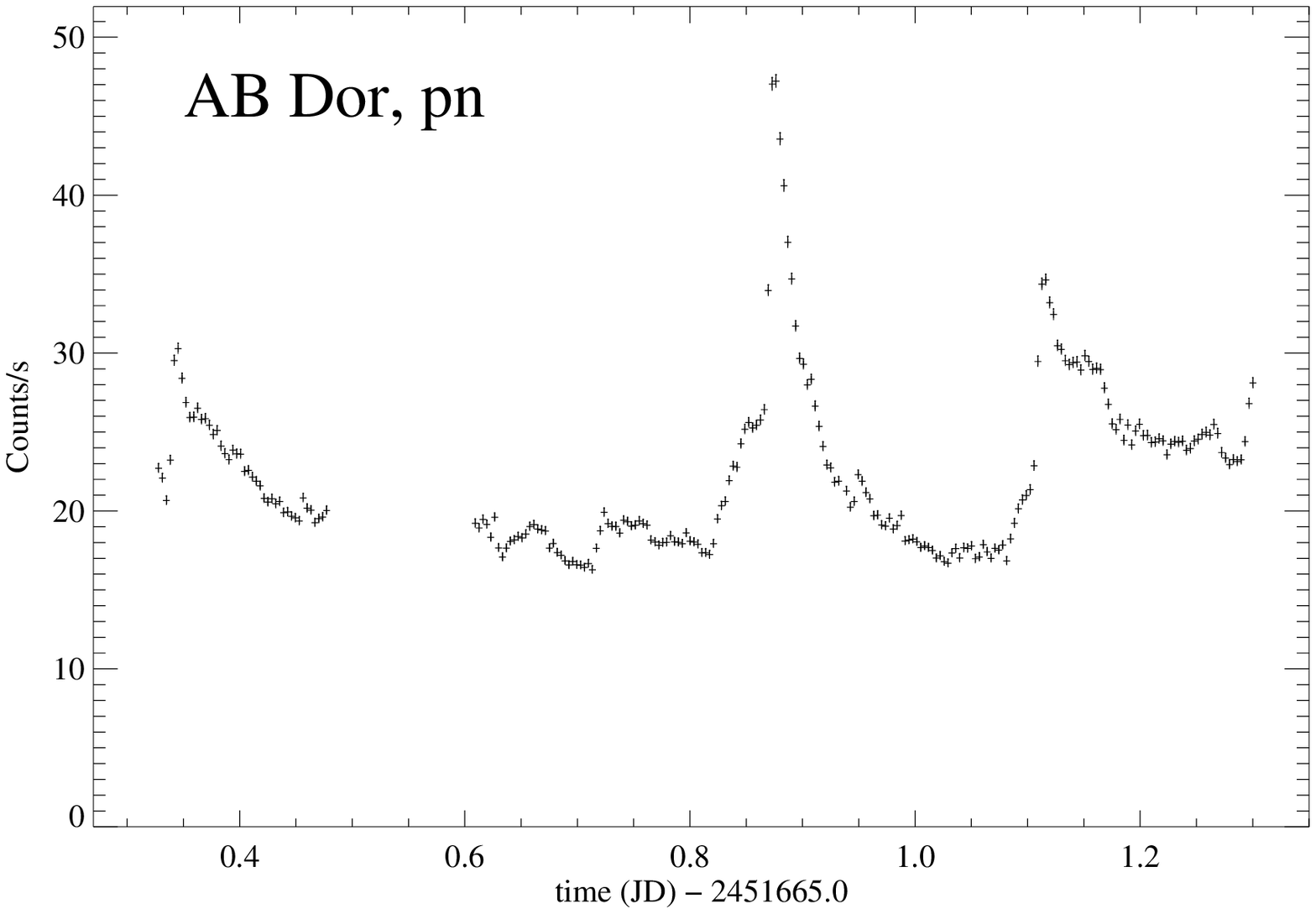}
\caption{{\it Left}: Total 1st and 2nd order RGS light curve of HR 1099. {\it Right}: 
EPIC pn light curve of AB Dor (April 30/May 1).}
\end{figure}
%___________________________________________________________________________
\begin{figure}[!t]
  \plotfiddle{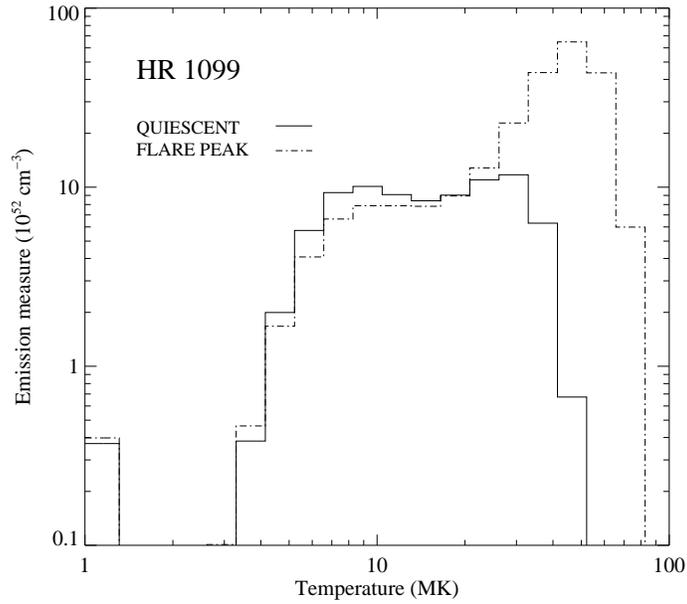}{7.25cm}{0}{45}{45}{-130}{-10}
  \caption{Realisations of the quiescent (solid) and flare peak (dot-dash) EM 
   distributions for HR 1099 using Chebychev polynomials.}
\end{figure}
\begin{figure}[!b]
  \centering \leavevmode
  \epsfxsize=0.96\textwidth \epsfbox{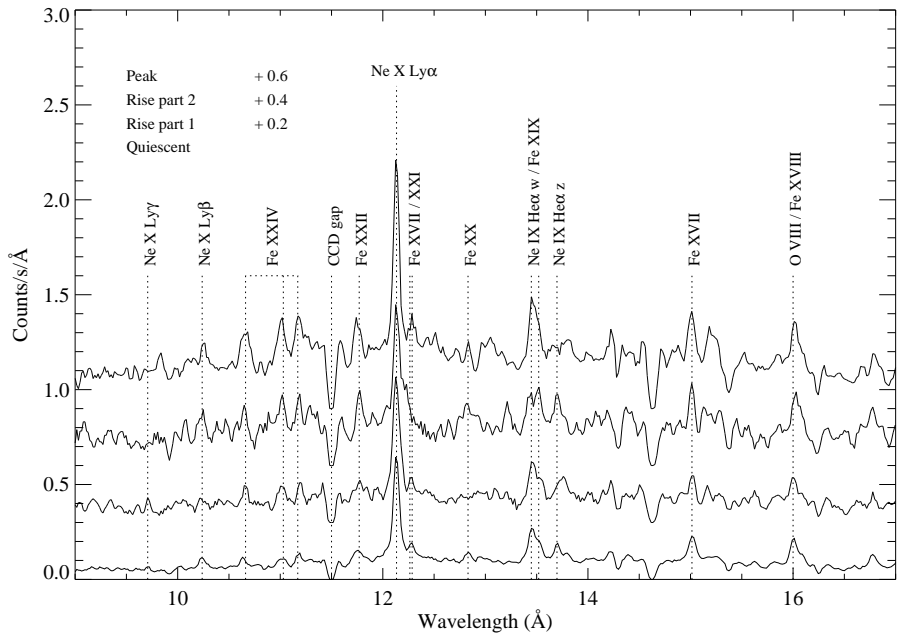}
  \caption{Extract from smoothed RGS spectra of the four time segments (lowest:
   quiescent; uppermost: flare peak). Note relative shifts of
   the spectra along the vertical axis.}
  \end{figure}
%___________________________________________________________________________
\section{RGS spectra of flares}
\begin{figure}[!t]
  \centering \leavevmode
  \epsfxsize=0.80\textwidth \epsfbox{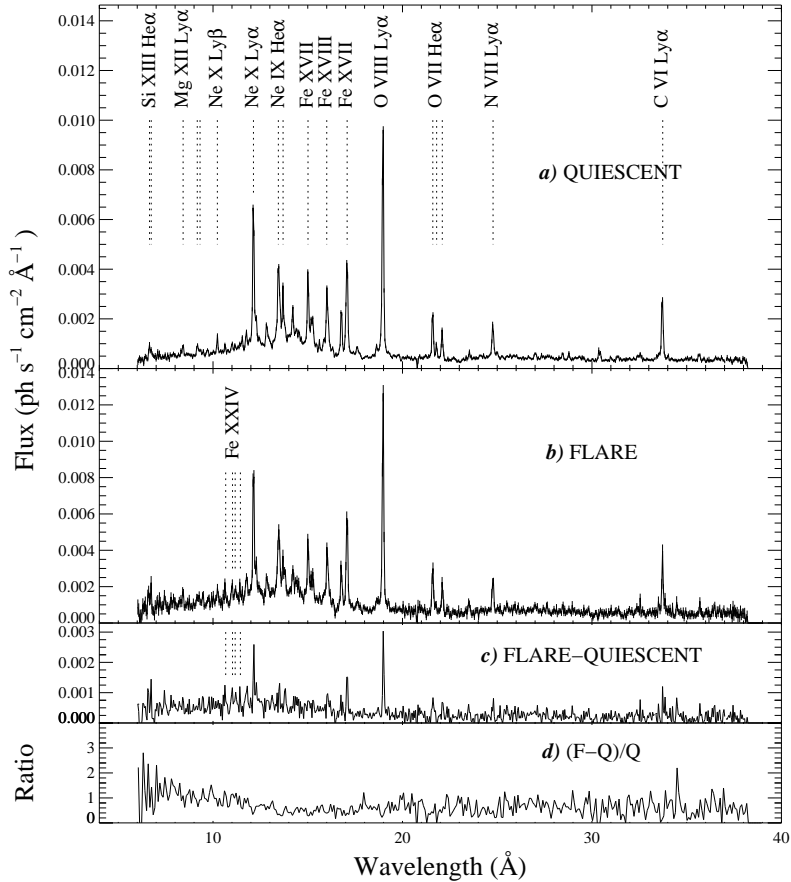}
  \caption{RGS fluxed spectrum of the {\bf (a)} quiescent and {\bf (b)} 
    flaring AB Dor, {\bf (c)} the difference spectrum, and {\bf (d)} the ratio
    ``(flare$-$quiescent)/quiescent''. Data are binned to a resolution of 
    0.04375~\AA\ for (a-c) and to 0.0875~\AA\ for (d). Note different flux 
    scales. The Fe~{\sc xxiv} lines and an excess continuum shortward of 
    10~\AA\ are evident in  (c) and (d).}
\end{figure}
Time-dependent spectroscopy has been performed on the flare data sets (HR 1099,
AB Dor, YY Gem).
We have first studied the quiescent emission measure (EM) distributions and found
them to be continuous over a wide range of temperatures (e.g., Fig.~2). The EM
of HR 1099, for example, drops to very low
values below $\approx$5~MK. A small but significant EM is needed around 1--3~MK in order to
explain ``cool'' lines such as the bright C~{\sc vi}~Ly$\alpha$ line and the
O~{\sc vii} triplet around 22~\AA. Audard et al.~(2001) further constrain the EM
of Capella at the coolest temperatures using dielectronic recombination satellite lines of
O~{\sc vi}. During the flare on HR 1099, the EM changes considerably but only at
high temperatures, while the cooler plasma remains essentially unchanged,
indicating that the flare does not affect a major part of the corona (Fig.~2). 
Above 30~MK, however, a very significant EM component develops, with contributions
from plasma up to about 90~MK (Fig.~2). During this strong heating, Fe~{\sc xxiv} lines
become significantly enhanced (Fig.~3). The rapid flares on AB Dor (Fig.~1,
right) show similar
spectral features: strong Fe~{\sc xxiv} lines become apparent during the 
flares, and the continuum becomes enhanced shortward of 12~\AA\ (Fig.~4).
%___________________________________________________________________________
\section{Variable abundances}
Elemental abundances vary significantly during the observed flares. 
The metal abundance during one of AB
Dor's flare increases by a factor of $\approx$3 and decays back to its
quiescent level (G\"udel et al.~2001a). In the HR 1099 flare spectra, the Fe and Si elemental
abundances increase by factors of $\approx$3, and 10 (despite large error bars,
the latter abundance increases by a factor of 6 at least), respectively, 
while the Ne (high-FIP) abundance does \emph{not} increase, staying constant 
at its quiescent level, within the confidence ranges (Audard et al.~2001).
%___________________________________________________________________________
\section{Densities}
\begin{figure}
\plotfiddle{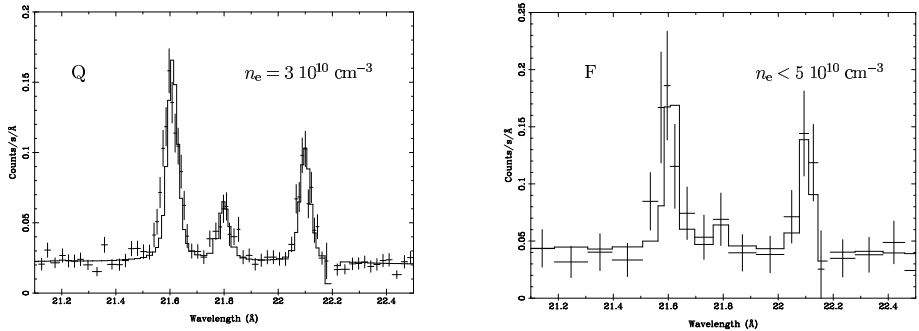}{4cm}{-90}{55}{55}{-210}{150}
\caption{Density-sensitive line triplet of He-like O~{\sc vii} (resonance,
intercombination, and forbidden line for increasing wavelength).
Left panel shows best-fit (histogram) to RGS data (error bars) during AB Dor's 
quiescence, while right panel shows the triplet for data selected during 
flares.}
\end{figure}
The He-like O~{\sc vii} triplet around 22~\AA\ provides insight into the
\emph{average} density of the ``cool'' plasma component (around 1--3~MK).
Significant quiescent densities have been found in AB Dor ($n_{\rm e} = [3 \pm
1.5] \times 10^{10}$~cm$^{-3}$; G\"udel et al.~2001a), YY Gem ($n_{\rm e} =
1.6^{+2.6}_{-1.4} \times 10^{10}$~cm$^{-3}$; G\"udel et al.~2001b), while
only upper limits can be derived for HR 1099 ($<1\times 10^{10}$~cm$^{-3}$;
Audard et al.~2001) and Castor ($<3\times 10^{10}$~cm$^{-3}$; G\"udel et al.
2001b). No appreciable change in the \emph{average} density of the emitting 
\emph{cooler} material has been found during flares (Fig.~5). Recalling Fig.~2, 
it implies that the flares predominantly affect hot material and have limited 
influence on the EM of the relatively cool plasma.
%___________________________________________________________________________
\section{Summary and Conclusions}
Flares are frequent in the stellar coronal sources
observed so far by \textit{XMM-Newton}. Large flares occurred on HR 1099, AB Dor, and YY
Gem, allowing us to perform time-dependent spectroscopy with the high-resolution
RGS in the range 5--35~\AA\  (0.35--2.5~keV).
The presence of very hot (up to several tens of MK) material has been inferred,
and evidence for elemental abundance enhancements has been found.
Low-FIP elements appear to increase significantly during a flare on HR 1099, while
the high-FIP element Ne stays at constant abundance. This behaviour contrasts
with the \emph{inverse} FIP effect found in these coronae during quiescence. 
Our \textit{XMM-Newton} observations thus indicate that flares are
also important as agents that relate chromospheric/photospheric plasma with
coronal plasma. 

The flare EM distributions appear to be composed of a very hot plasma 
(up to $100$~MK) that evolves  rapidly, and of a stable quiescent plasma. This
result is consistent with the absence of a change in the density of the cool
material. Stellar flares are important probes for stellar coronal plasma and the coronal
heating mechanism. The very high temperatures attained by these flares are unknown
to solar flares. Such observations are important to extend solar knowledge to
more extreme conditions appropriate for magnetically active stars.

\acknowledgements
We thank all XMM teams for their support in the evaluation of the present and
related data. The PSI group is supported
by the Swiss Academy of Natural Sciences and the Swiss National Science
Foundation (grants 2100-049343 and 2000-058827). SRON is financially
supported by the Netherlands Organization for Scientific Research (NWO).
The Columbia University team acknowledges generous support from the
National Aeronautics and Space Administration.  MSSL acknowledges support 
from the Particle Physics and
Astronomy Research Council. This work is based on observations obtained with
XMM-Newton, an ESA science mission with instruments and contributions directly
funded by ESA member states and the USA (NASA).

\end{document}